\documentclass[twocolumn,showpacs,amsmath,amstex,amssymb,mathfonts,prb]{revtex4-1}
\usepackage{amsthm,amsfonts,graphicx,verbatim}
\usepackage{amsmath}
\usepackage{amssymb}
\usepackage{amsthm}
\usepackage{amsfonts}
\usepackage{listings}
\usepackage{enumerate}
\usepackage{latexsym}
\usepackage{psfrag}
\usepackage{bm}
\usepackage{graphicx}
\usepackage{subfigure}
\newcommand{\be}{\begin{equation}}
\newcommand{\ee}{\end{equation}}
\newcommand{\bea}{\begin{eqnarray}}
\newcommand{\eea}{\end{eqnarray}}

\renewcommand{\phi}{\varphi}
\renewcommand{\epsilon}{\varepsilon}

\begin{document}

\title{Tunable interactions and phase transitions in Dirac materials in a magnetic field}

\author{Z. Papi\'c$^1$, D. A. Abanin$^2$, Y. Barlas$^3$, and R. N. Bhatt$^{1,2}$}
\affiliation{$^1$ Department of Electrical Engineering, Princeton University, Princeton, NJ 08544}
\affiliation{$^2$ Princeton
 Center for Theoretical Science, Princeton University, Princeton, NJ
 08544} 
\affiliation{$^3$ National High Magnetic Field Laboratory and Department of Physics, Florida State University, FL 32306, USA}


\date{\today}

\begin{abstract}

A partially filled Landau level (LL) hosts a variety of correlated states of matter with unique properties. The ability to control these phases requires tuning the effective electron interactions within a LL, which has been difficult to achieve in GaAs-based structures. Here we consider a class of Dirac materials in which the chiral band structure, along with the mass term, give rise to a wide tunability of the effective interactions by the magnetic field. This tunability is such that different phases can occur in a \emph{single} LL, and phase transitions between them can be driven \emph{in situ}. The incompressible, Abelian and non-Abelian, liquids
are stabilized in new interaction regimes. Our study points to a realistic method of controlling the correlated phases and studying the phase transitions between them in materials such as graphene, bilayer graphene, and topological insulators. 

\end{abstract}
\maketitle

In a magnetic field, the kinetic energy of electrons confined to move in two dimensions is quenched into a set of discrete Landau levels (LLs). The properties of a partially filled LL are therefore determined solely by the electron-electron interactions, which give rise to a number of fundamentally new many-body phases, including the incompressible fractional quantum Hall (FQH) states~\cite{tsg, laughlin, prange, jainbook}, compressible Fermi-liquid-like states (CFL)~\cite{HLR}, as well as states with broken translational symmetry, such as charge-density waves (CDW), stripes and bubble phases~\cite{stripe_bubble}.    

The competition between different phases at a given partial filling $\nu$ is sensitive to the form of the effective Coulomb interactions within a LL. Experimentally, to date there is no reliable way of controlling the effective interactions within a LL~\cite{tilted_wqw}. This is because in GaAs, the most common high-mobility 2D electron system, the effective interactions depend only on the LL number, and not on the magnetic field. Thus, in order to control the FQH phases, it is advantageous to find alternative 2D electron systems with tunable interactions~\cite{tunable}.

Recently, a new class of such materials clean enough to observe FQH phases has emerged. These so-called Dirac materials host chiral excitations with the non-trivial Berry phases. Examples include graphene, bilayer graphene~\cite{CastroNeto09}, topological insulators~\cite{KaneHasan}, as well as certain quantum wells~\cite{HgTe}. A natural question arises -- could the Dirac materials offer any tunability in the FQH regime? 

Here, we answer this question affirmatively, and provide a realistic model where the effective interactions can be widely tuned by varying an external magnetic field, giving rise to several phases within {\it each} LL, and quantum phase transitions (QPTs) between them.  We map out the phase diagram of chiral materials at certain filling factors, identifying a new regime of the effective interactions where the non-Abelian Moore-Read (MR)~\cite{Moore91} and other paired states~\cite{ReadRezayi} are stable away from $n=1$ LL of GaAs~\cite{Morf98,Rezayi00,ReadRezayi}. Finally, we predict several types of phase transitions that occur in Dirac materials (from an incompressible FQH state to a compressible state, with or without breaking of translational symmetry; between an Abelian and a non-Abelian FQH state), which previously have only been considered in artificial theoretical models. 

The control of the correlated phases proposed here is attractive for two reasons. First, it allows one to realize and stabilize the exotic states~\cite{ReadRezayi}, or may lead to a discovery of new correlated states that are weak or absent in GaAs. Second, it provides a setting for studying the fundamental problem of phase transitions that involve topologically ordered states. In our model, both of these goals can be achieved by current experimental techniques of tuning the gap in the spectrum, e.g. in bilayer graphene~\cite{CastroNeto09}, and topological insulators~\cite{KaneHasan}. In the former case, the gap is opened by a perpendicular electric field, and in the latter case by the deposition of magnetic adatoms. In graphene, it is more challenging to open the gap, although there exist several promising proposals (e.g., the mass can be generated either spontaneously, or as a result of sublattice symmetry breaking~\cite{CastroNeto09}).

We consider fermions with the Berry phase $\pi$ and $2\pi$, described by the $2\times 2$ Hamiltonian, 
\be\label{eq:hamiltonian}
H_{\lambda\pi}=
 \left[\begin{array}{cc}
        \Delta &  \mathcal{M}_\lambda(p_x+ip_y)^\lambda\\
         \mathcal{M}_\lambda(p_x-ip_y)^\lambda & -\Delta
     \end{array}
 \right] , \; \lambda=1,2
\ee
where $\mathcal{M}_1 \equiv v_0$ is the Fermi velocity, $\mathcal{M}_2 \equiv 1/2m$ ($m$ is the effective mass), and $2\Delta$ is the band gap. The case of $\pi$-carriers ($\lambda=1$) is realized in graphene, topological insulators, and special quantum wells~\cite{HgTe}; the case of $2\pi$-carriers ($\lambda=2$) occurs in bilayer graphene~\cite{CastroNeto09}.

Landau level spectrum for $\pi$-carriers, obtained by solving the Schr\"{o}dinger equation for $H_{\pi}$ in a magnetic field, is given by $\epsilon_n={\rm sgn}(n)\sqrt{\Delta^2+\epsilon_0^2 |n|}$, $n=\pm 1, \pm 2, ...$,  $\epsilon_0=\sqrt{2} \hbar v_0/\ell_B$ is the characteristic energy scale, and $\ell_B=\sqrt{\hbar c/eB}$ is the magnetic length. The corresponding two-component wave functions are given by $\psi_n=(\cos\theta_n \phi_{|n|-1},\sin\theta_n \phi_{|n|})$, where $\phi_n$ is the wave function of the $n$th non-relativistic LL (standard magnetic oscillator wave function), and parameter $\theta$ can be expressed as $\tan\theta_n=\left[ {\rm sgn}(n)\sqrt{(\Delta/\epsilon_0)^2+|n|}-\Delta/\epsilon_0 \right]/\sqrt{|n|}$.
In the limit of zero mass (graphene case), $\tan\theta_n=\pm 1$, and the weights of $\phi_{|n|-1}$ and $\phi_{|n|}$ in the wave function $\psi_n$ become equal. In the opposite limit of very large mass, $\Delta/\epsilon_0\gg 1$, the LLs become identical to the non-relativistic ones: $\tan\theta_{|n|}\to 0$ (this corresponds to 
$\psi_{|n|}\approx (\phi_{|n|-1} ,0)$), and $\tan\theta_{-|n|}\to -\infty $ ($\psi_{-|n|}\approx (0,\phi_{|n|})$). As we comment below, varying $\Delta/\epsilon_0$ between $0$ and $\infty$, which can be achieved by changing magnetic field, allows one to explore the whole range $\theta\in (0;\pi/2)$.  

\begin{figure}[ttt]
\includegraphics[scale=0.4]{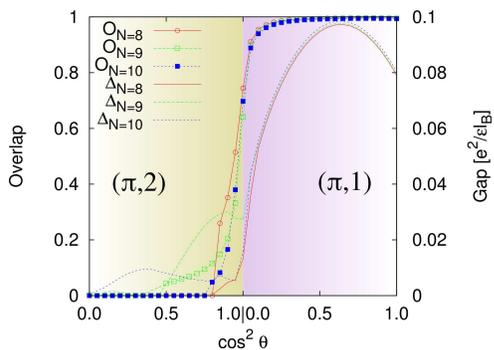}
\vspace{-1mm}
\caption[]{(Color online) Transition between the Laughlin and the bubble state. Overlap with the Laughlin wavefunction $\mathcal{O}_N$ and charge gap $\Delta_N$ are plotted for an $N$-particle system at $\nu=1/3$ filling of $(\pi,1)$ and $(\pi,2)$ LLs.}
\label{fig:laughlin}
\vspace{-0pt}
\end{figure}
The two-component nature of the wavefunction modifies the effective interaction within a LL, which determines the many-body phases at a partial filling. We use the standard approximation and project the interaction onto a partially filled LL. In this case, the Fourier transform of the effective interaction is a product of bare Coulomb interaction, $V(q)=2\pi e^2/q$, and the form factor~\cite{prange} $|F_n(q)|^2$ which contains the information about the band structure. In case of the materials with the Berry phase $\pi$,
\be\label{eq:formfactors}
F_{n}^{\pi} (q)=\cos^2\theta L_{|n|-1} (q^2 \ell_B^2/2)+\sin^2\theta L_{|n|}(q^2 \ell_B^2 /2),
\ee
where $L_{k}$ is the $k$th Laguerre polynomial, and for simplicity we omitted the index of $\theta$. The form-factor is a mixture of the $(|n|-1)$th and $|n|$th LL form-factors in a non-relativistic 2DES with parabolic dispersion. At  $\theta=\pi/4$, the above equation reduces to the form-factor of pristine graphene~\cite{Nomura06}. 

Similarly, for carriers with the Berry phase $2\pi$, the LL spectrum is given by $\epsilon_n={\rm sgn}(n) \sqrt{\Delta^2+ \epsilon_c^2 |n|(|n|+1)}$, $n=\pm 1,\pm 2, ...$, where $\epsilon_c=eB/mc$ is the cyclotron energy. The corresponding wave functions are $\psi_n=(\cos\theta_n \phi_{|n|-1}, \sin\theta_n \phi_{|n|+1} )$, with $\tan\theta_n= \left[ {\rm sgn}(n)\sqrt{(\Delta/\epsilon_c)^2+|n|(|n|+1)}-\Delta/\epsilon_c \right] / \sqrt{|n|(|n|+1)}$. The form-factor is a mixture of standard $(|n|-1)$th and $(|n|+1)$th form-factors, 
\be\label{eq:formfactors2}
F_{n}^{2\pi} (q)=\cos^2\theta L_{|n|-1} (q^2 \ell_B^2/2)+\sin^2\theta L_{|n|+1}(q^2 \ell_B^2 /2). 
\ee
The tunable form of the effective interactions Eqs.(\ref{eq:formfactors},\ref{eq:formfactors2}) provides a way to analyze the transitions between strongly correlated phases in a direct manner, as will be demonstrated below. 

Much of the previous theoretical work on the quantum Hall effect in the Dirac materials has been limited to graphene, exploring the consequences of the four-fold LL degeneracy (valley and spin) that leads to new SU(2) and SU(4)-symmetric fractional states~\cite{multi-component}. Here we neglect the multicomponent degrees of freedom~\cite{footnote}, and examine the effects originating from the interplay of the Coulomb interaction and band structure. The large variation of the effective interactions, due to the band structure, is assumed to be the dominant effect, even when corrections due to LL mixing~\cite{llmixing} are taken into account. 

\begin{figure}[ttt]
\includegraphics[scale=0.4]{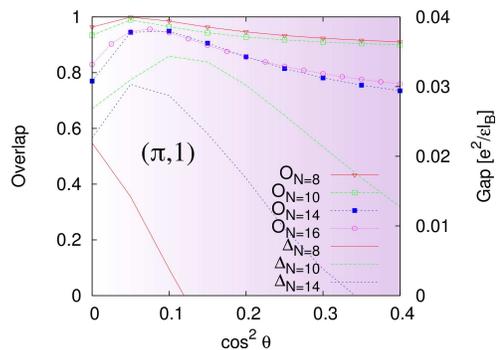}
\vspace{-1mm}
\caption[]{(Color online) Transition between the Moore-Read and the CFL state. Overlap with the Pfaffian wavefunction $\mathcal{O}_N$ and charge gap $\Delta_N$ are plotted for an $N$-particle system at $\nu=1/2$ filling of $(\pi,1)$ LL.}
\label{fig:pf}
\vspace{-0pt}
\end{figure}
\begin{figure*}[t]
  \begin{minipage}[l]{\linewidth}
    \includegraphics[scale=0.85]{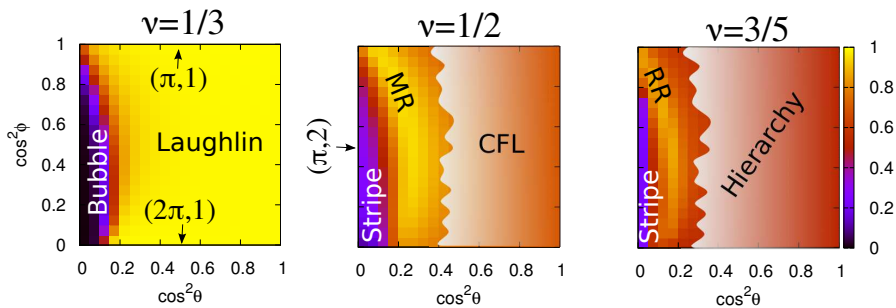}
  \end{minipage}
\caption{(Color online) Phase diagram of a generic chiral material in a partially filled LL containing a mixture of $n=0$, $n=1$ and $n=2$ non-relativistic form-factors. Color scheme on the right defines the magnitude of the overlaps with the Laughlin, Moore-Read and the $k=3$ Read-Rezayi (RR) wavefunction. Shaded regions represent the phases with different topological numbers (CFL, hierarchy state). The phase boundary between topological states is approximately drawn by a wiggly line.}
\label{fig:phasediagram}
\vspace{-0pt}
\end{figure*}
We proceed by studying the interacting states of Dirac materials using exact diagonalization in the spherical and torus geometry~\cite{Haldane86, pbc}. The former is useful in studying the incompressible liquids, but is not suitable for states that break translational symmetry, when periodic boundary conditions are more natural~\cite{pbc}. If the system has an underlying lattice structure, it is assumed that its lattice constant is much smaller than $\ell_B$. We map out the phase diagram as a function of $\theta$ using the overlaps between an exact ground state and a trial wavefunction. Moreover, we evaluate the energy gaps for creating charged excitations, $\Delta\equiv E_{qp}+E_{qh}-2E_0$, where $E_{qh} (E_{qp})$ is the energy of the system in the presence of a quasihole (quasiparticle) and $E_0$ is the ground state energy (all in units of $e^2/\epsilon \ell_B$). For simplicity, when we use the spherical geometry, the form factors Eqs.(\ref{eq:formfactors}, \ref{eq:formfactors2}) are calculated for a flat surface; the curvature is expected to produce only small quantitative difference in larger systems and has a stronger impact on the energy gaps than the overlaps.  Various choices of the single-particle Hamiltonian are denoted by $(\lambda\pi,n)$, where we restrict to $\lambda=1,2$ and $n=1,2$. Higher values of $n$ are not considered because they do not support FQH states.

At partial filling $\nu=1/3$, the Laughlin state is robust in $(\pi,1)$, $(\pi,2)$, and $(2\pi,1)$ LLs. The overlaps between the exact ground state and the Laughlin wavefunction, as well as the charge gaps in $(\pi,1)$ and $(\pi,2)$ LLs, are illustrated in Fig.~\ref{fig:laughlin}. The overlap remains very close to one for $\cos^2\theta\in \left[ 0.2;1\right]$ in the $(\pi,1)$ LL. This is not surprising because the form-factors of this LL are a mixture of $0$ and $1$ non-relativistic form-factors, both of which favor the Laughlin state. An interesting feature of Fig.~\ref{fig:laughlin} is that the energy gap shows a maximum at $\cos^2\theta \approx 0.65$, rather than at $\cos^2\theta=1$ (pure $n=0$ non-relativistic LL). This special value of $\theta$ maximizes the ratio of $V_1/V_3$ Haldane pseudopotentials~\cite{prange} which controls the gap in this case. The overlap and the gap decrease drastically as $\cos^2\theta \to 0$; furthermore, in $(\pi,2)$ LL, the Laughlin state quickly undergoes a QPT to a bubble phase as the overlaps and gaps drop to zero. 
A signature of this transition is also detected in the projected structure factor~\cite{prange}, which develops a sharp peak in the bubble phase~\cite{unpublished}. The transition is more naturally captured in the torus geometry, where a large ground state degeneracy~\cite{stripe_bubble_numerics}, characterized by a 2D array of crystal momenta, develops in  $(\pi,2)$ LL.  Similar behavior is found in $(2\pi,1)$ LL~\cite{unpublished}. One of the of the members of the multiplet belongs to the zero-momentum sector, which suggests that the transition is likely second-order.

Next, we consider a half-filled LL, where we find evidence for the Moore-Read~\cite{Moore91} correlations in half-filled chiral $(\pi,1)$ and $(2\pi,1)$ LLs. In the former case, the Pfaffian becomes more robust than its GaAs analogue for $\cos^2 \theta \approx 0.1$, Fig.~\ref{fig:pf}. The point $\cos\theta=0$ corresponds to a $n=1$ non-relativistic LL where the overlap is high (as expected). As $\cos\theta$ increases, the overlap reaches a maximum value at $\cos^2\theta\approx 0.1$; this is followed by an enhancement of the gap, which peaks at approximately the same point (gaps show stronger finite-size effects than the overlaps). The increased stability of the Pfaffian was previously discussed in Ref.~\cite{Apalkov11} in the context of biased bilayer graphene. On the other hand, away from the ``optimal" point $\cos^2\theta\approx 0.1$, the Pfaffian  undergoes a QPT to the compressible CFL state~\cite{HLR, Rezayi00}. This transition is accompanied by a change in \emph{shift}~\cite{shift} of the ground state beyond $\cos^2 \theta=0.4$, and is also manifested in the gaps dropping to zero (Fig.\ref{fig:pf}). A similar scenario is found in $(2\pi,1)$ LL~\cite{unpublished}.
At larger values of $\cos^2\theta$, in both $(\pi,1)$ and $(2\pi,1)$,  we expect a QPT into the CFL state; at very small $\cos^2\theta$ in $(2\pi,1)$ LL, the Pfaffian gives way to a stripe phase. Thus, chiral materials are suitable for studying phase transitions between the Pfaffian, stripe phase and the CFL; these transitions are expected to be of different nature, either first or second order~\cite{Bonesteel}. This is a distinct advantage over GaAs-based 2DES where the effective interactions are significantly more difficult to tune in a controlled fashion. 

Note that the enhancement of the overlaps and gaps might appear similar in nature to tweaking of the $V_1$ pseudopotential, known to have a favorable effect on the Pfaffian state~\cite{Rezayi00}. However, in the present case, the interaction that favors the MR state is a superposition of $n=0$ and $n=1$ form factors, including the long range tail of the repulsive potential; thus it represents a new regime where the MR state is stable. This new regime of stability does not crucially depend on the presence of $n=1$ LL form-factor. Similar signatures of the paired states occur also in $(2\pi,1)$ LL, where the interaction involves a superposition of $n=0$ and $n=2$ LL form-factors, making the difference from GaAs even more striking. 

We have also analyzed the filling $\nu=3/5$~\cite{unpublished} (or $\nu=2/5$), where the one expects to find the non-Abelian $k=3$ Read-Rezayi state~\cite{ReadRezayi} that supports universal topological quantum computation~\cite{tqc}. Similar to the MR state, in chiral materials we find non-Abelian correlations in both $(\pi,1)$ and $(2\pi,1)$ LLs, with a phase transition to the stripe phase and the Abelian hierarchy state~\cite{Haldane86} (see also Fig.\ref{fig:phasediagram}). 

Finally, we consider a model that generalizes the effective interaction given above and allows to map the phase diagram for a wider class of materials. Specifically, we study the form-factors that are represented as a linear combination of $|n|$, $|n|+1$ and $|n|+2$ non-relativistic form-factors. Such form-factors define a two-parameter family, 
\bea\label{eq:formfactor_general}
\nonumber F_n(q)&=& \cos^2\theta L_{|n|} (q^2 \ell_B^2/2)  +\sin^2\theta \cos^2\phi L_{|n|+1}(q^2 \ell_B^2 /2) \\  
 && +\sin^2\theta \sin^2\phi L_{|n|+2}(q^2 \ell_B^2 /2). 
\eea
The effective interactions of the above form arise in a number of materials, including trilayer graphene~\cite{trilayer}, as well as bilayer graphene in the limit of large asymmetry between the two layers~\cite{McCann06,Apalkov11}. The exact relation of the parameters $(\theta,\phi)$ to the band structure of various materials will be discussed elsewhere~\cite{unpublished}.  
The phase diagram of the model in Eq.(\ref{eq:formfactor_general}) for its $n=0$ LL is illustrated in Fig.~\ref{fig:phasediagram}. Along certain axes (indicated by arrows), the generalized model reduces to one of the particular cases $(\lambda\pi,n)$ presented earlier. A salient feature of the generalized model is that non-Abelian states are found in a strip of $\theta, \phi$ values where the effective interaction significantly deviates from $n=1$ non-relativistic one. On the right side of the strip, the states undergo a QPT to another FQH (or CFL) state with different topological numbers. These transitions are represented by wiggly lines in Fig. \ref{fig:phasediagram}, where the shaded areas mark the resulting phases with a different shift. On the left side, the system crosses over to compressible, CDW-like phases. The Abelian states dominate over a wide region of parameter space and are insensitive to variation in $\phi$, unless $\cos\theta$ is close to zero. We note that the presented phase diagram is consistent with the one obtained by considering the charge gaps instead of the overlaps~\cite{unpublished}. 

In summary, we showed that the new regimes of the effective Coulomb interactions can be realized in chiral materials. This allows one to stabilize the desired phases (including non-Abelian ones) within a single LL, and provides a way to engineer QPTs between them. Our results apply to a number of available high-mobility 2DES~\cite{CastroNeto09, trilayer, KaneHasan}, however bilayer and trilayer graphene appear as most suitable candidates because their band structure can be tuned by an external electric field. In fact, our proposal can be already be realized in bilayer graphene with the existing experimental tools. It was demonstrated~\cite{yzhang} that the gap can be controlled in the interval $\pm 125{\rm meV}$. With the cyclotron energy of $\sim 20{\rm meV}$ at $B=10{\rm T}$ this translates into the ability to tune $\cos^2 \theta$ in the interval $(0.01; 0.99)$, sufficient for the realization of the non-Abelian states (see Fig.~\ref{fig:phasediagram}). We believe that experimental observation of the phases and QPTs predicted here should be feasible in the near future, as the fractional quantum Hall states in graphene have already been observed~\cite{graphene_fqhe}, and transport anomalies suggesting fractional states have been seen in topological insulators~\cite{Phuan11} and bilayer graphene~\cite{Lau10}.

{\sl Acknowledgements}. This work was supported by DOE grant DE-SC$0002140$. Y.B. was supported by the State of Florida.

\bibliography{paper}

\begin{thebibliography}{99}
\bibitem{footnote}
The presence of a mass and spin Zeeman term will lead to valley and spin broken symmetry FQH states with exchange enhanced gaps.
\end{thebibliography}

\end{document}